\documentclass[prd,aps,twocolumn,nofootinbib,superscriptaddress]{revtex4-1}

\usepackage{graphicx}
\usepackage{xcolor,amsmath,amssymb}
\usepackage{multirow}
\usepackage{comment}
\usepackage{enumitem}
\usepackage{hyperref}
\hypersetup{colorlinks, linkcolor={teal},citecolor={blue},urlcolor={blue}}

\begin{document}

\title{Distance duality relation in symmetric teleparallel gravity}

\author{Rocco D'Agostino}
\email{rocco.dagostino@inaf.it}
\affiliation{INAF -- Osservatorio Astronomico di Roma, Via Frascati 33, 00078 Monte Porzio Catone, Italy}
\affiliation{INFN, Sezione di Roma 1, Piazzale Aldo Moro 2, 00185 Roma, Italy}

\begin{abstract}
In this work, we investigate the distance duality relation (DDR) in symmetric teleparallel theories, where gravity is mediated by nonmetricity.
Starting from the general metric-affine formulation and adopting the geometrical optics approximation, we show that the standard Etherington reciprocity relation remains valid in the presence of nonmetricity when electromagnetism is minimally coupled and the photon number is conserved.
We then extend the analysis to a class of $f(Q)$ theories with a nonminimal coupling between the electromagnetic field and the nonmetricity scalar. We demonstrate that such an interaction modifies the conservation of the photon number current, leading to a dynamical violation of the DDR. 
Focusing on a homogeneous and isotropic spacetime background in the coincident gauge, we derive a generalized DDR formula that directly relates observational distance measures to the Hubble expansion rate. 
Furthermore, we discuss the link between the deviations from Etherington's relation and variations of the effective fine-structure constant.
Specific illustrative examples of the coupling function are also analyzed, showing that phenomenologically viable models predict only small deviations from the standard DDR.
Our results provide a unified framework to distinguish between the geometric and dynamical origins of DDR violations, opening new avenues for testing non-Riemannian gravity with future high-precision astrophysical and cosmological observations.
\end{abstract}

\maketitle

\section{Introduction}

The success of general relativity (GR) in describing gravitational phenomena from the solar system to the scale of the Universe is limited by a number of observational and theoretical challenges \cite{Berti:2015itd,Debono:2016vkp,Ishak:2018his}. In particular, the accelerated expansion of the Universe, the dynamics of galaxies and clusters, and the apparent presence of dark matter and dark energy point to phenomena that are difficult to explain within the standard GR framework without introducing unknown components of the cosmic energy budget \cite{Weinberg:1988cp,Padmanabhan:2002ji,Peebles:2002gy,Huterer:2017buf,DAgostino:2019wko,Arbey:2021gdg,DAgostino:2022fcx}.
Furthermore, the standard $\Lambda$CDM cosmological model is affected by persistent tensions between independent observational probes, most notably the discrepancy in the value of the Hubble constant inferred from early and late time measurements and the mismatch in the amplitude of matter fluctuations between cosmic microwave background data and large-scale structure observations \cite{Riess:2019cxk,DAgostino:2023cgx,CosmoVerseNetwork:2025alb,DESI:2025zgx}. These anomalies may originate from unresolved systematics, but they may also hint at new physics beyond $\Lambda$CDM \cite{Joyce:2014kja,Perivolaropoulos:2021jda,DAgostino:2023tgm,Califano:2024tns}.

These challenges suggest that GR may not provide a complete description of gravity at all scales. Over the past two decades, this has motivated the development of modified theories of gravity, which extend or alter the geometric and dynamical foundations of GR \cite{Clifton:2011jh,Koyama:2015vza,Nojiri:2017ncd,Capozziello:2019cav,DAgostino:2020dhv}. Such theories aim to reproduce the successes of GR on well-tested scales while providing alternative explanations for cosmological observations. Examples include extended gravitational Lagrangians where the Ricci scalar is generalized to a function \cite{Carroll:2003wy,Sotiriou:2008rp,Nojiri:2010wj,DeFelice:2010aj,DAgostino:2024sgm,DAgostino:2025axy},
scalar-tensor theories, where gravity is mediated by both a metric tensor and one or more scalar fields \cite{Brans:1961sx,Horndeski:1974wa,Elizalde:2004mq,Ezquiaga:2017ekz,DAgostino:2019hvh,Califano:2023aji}, and nonlocal theories, which incorporate interactions depending on the global properties of spacetime rather than only local curvature \cite{Deser:2007jk,Calcagni:2007ru,Hehl:2008eu,Biswas:2011ar,Maggiore:2014sia,Capozziello:2022rac,DAgostino:2025wgl,DAgostino:2025sta,DAgostino:2026wln,DAgostino:2026hht}. 
In addition, teleparallel theories have been extensively studied, in which gravity is encoded in torsion rather than curvature \cite{Hehl:1976kj,Bengochea:2008gz,Linder:2010py,Geng:2011aj,Cai:2015emx,Abedi:2018lkr,DAgostino:2018ngy,DAgostino:2025kme}. More generally, non-Riemannian frameworks explore spacetime geometries in which nonmetricity can each play a fundamental role, offering new avenues to address cosmological and astrophysical phenomena beyond the reach of standard GR \cite{BeltranJimenez:2017tkd,Jarv:2018bgs,BeltranJimenez:2019tme,Hohmann:2021ast,DAmbrosio:2021pnd,Capozziello:2022wgl,DAgostino:2022tdk}.

A particularly powerful observational probe of modified gravity frameworks is the study of light propagation and the relations between different cosmological distances.
More specifically, the distance duality relation (DDR), first formulated by Etherington~\cite{Etherington:1933asu}, plays a foundational role in observational cosmology.
In GR, this relation follows from three key assumptions: (i) light propagates along null geodesics of a metric spacetime; (ii) the geodesic deviation equation governs the evolution of light beams; (iii) the number of photons is conserved along null congruences.

Deviations from the standard DDR, therefore, provide a window into the underlying nature of spacetime and gravitation \cite{Bassett:2003vu,Avgoustidis:2010ju,Hees:2014lfa,Holanda:2015zpz,Euclid:2020ojp,Matos:2023jkn,Jesus:2024nrl,DeLeo:2025rhy}.
Such deviations may arise from exotic photon interactions, opacity effects, or modifications of gravity. In particular, in non-Riemannian theories of gravity, the spacetime connection is no longer Levi-Civita, and the concepts of parallel transport, geodesics, and geodesic deviation are modified. Consequently, the assumptions underlying the standard derivation of Etherington's formula are no longer guaranteed to hold, making the DDR a sensitive probe of non-Riemannian spacetime geometries. Among these theories, symmetric teleparallel gravity (STG) and $f(Q)$ models stand out as particularly compelling frameworks, in which gravitational interactions are encoded purely in spacetime nonmetricity, while curvature and torsion vanish identically (see \cite{Heisenberg:2023lru} for a review).

The aim of this paper is twofold. 
First, we provide a systematic derivation of the DDR in symmetric teleparallel geometry, clarifying the role played by nonmetricity in the propagation of light beams and in the kinematical assumption underlying the Etherington relation.
Second, we extend the analysis to a dynamical $f(Q)$ theory with a nonminimal electromagnetic coupling, in which photon number conservation is explicitly broken. 

The introduction of a nonminimal coupling between the gravitational sector and the electromagnetic field is a well-explored avenue in modified gravity. Such interactions naturally arise, for instance, from one-loop quantum corrections in curved spacetime \cite{Drummond:1979pp}, and are often considered within an effective field theory approach. In this perspective, geometric scalar invariants can in principle couple to the electromagnetic operator $F_{\mu\nu}F^{\mu\nu}$, unless forbidden by specific symmetries \cite{Carroll:1998zi}. 
Couplings of this type have been widely studied in $f(R)$ and $f(T)$ theories to explain the evolution of cosmic magnetic fields and potential variations in the fine-structure constant \cite{Turner:1987bw,Bamba:2008ja,Bamba:2012mi}. Here, we extend this logic to the STG framework, using the nonmetricity scalar as the geometric mediator of the interaction.
This allows us to test the robustness of the DDR against purely geometric nonminimal interactions, providing a new perspective on how non-Riemannian degrees of freedom may affect cosmological observables.

The paper is organized as follows. In Sec.~\ref{sec:STG}, we review the fundamental principles of symmetric teleparallel gravity.
In Sec.~\ref{sec:metric-affine}, we analyze light propagation under minimal electromagnetic coupling, examining the kinematics of null geodesics and the validity of the distance duality relation in a nonmetric background. 
In Sec.~\ref{sec:f(Q)}, we extend our study to an $f(Q)$ theory with a nonminimal electromagnetic coupling. Here, we derive the transport equation for the electromagnetic amplitude and discuss the resulting dynamical violation of the DDR, providing an explicit relation for FLRW cosmology in the coincident gauge.
Finally, in Sec.~\ref{sec:final}, we summarize our main results and discuss their implications for future cosmological tests of STG gravity.

\section{Symmetric teleparallel gravity}
\label{sec:STG}

In a metric-affine spacetime, the fundamental geometrical objects are a
Lorentzian metric $g_{\mu\nu}$ and an independent affine connection
$\Gamma^\alpha{}_{\mu\nu}$.
The corresponding curvature, torsion and nonmetricity tensors are
defined, respectively, as \cite{BeltranJimenez:2017tkd}
\begin{align}
R^\alpha{}_{\beta\mu\nu} &=2\partial_{[\mu}\Gamma^\alpha{}_{\nu]\beta}+2\Gamma^\alpha{}_{[\mu|\lambda|}\Gamma^\lambda{}_{\nu]\beta}\,, \label{eq:R}\\
T^\alpha{}_{\mu\nu} &= 2\Gamma^\alpha{}_{[\mu\nu]}\,, \label{eq:T}\\
Q_{\alpha\mu\nu} &= \nabla_\alpha g_{\mu\nu}\,, \label{eq:Q}
\end{align}
where $\nabla_\alpha$ denotes the covariant derivative with respect to $\Gamma^\alpha{}_{\mu\nu}$\footnote{In Eqs.~\eqref{eq:R}--\eqref{eq:Q}, square brackets denote antisymmetrization of the enclosed indices, e.g., $A_{[\mu\nu]}=\frac{1}{2}(A_{\mu\nu}-A_{\nu\mu})$, while vertical bars indicate indices excluded from the antisymmetrization.}.

In STG, the connection is chosen such that
\begin{equation}
R^\alpha{}_{\beta\mu\nu} = 0\,, \quad
T^\alpha{}_{\mu\nu} = 0\,, \quad
Q_{\alpha\mu\nu} \neq 0 \,.
\end{equation}
Namely, curvature and torsion vanish identically, and gravity is entirely described by nonmetricity.
Under these constraints, the affine connection can always be decomposed as \cite{Nester:1998mp}
\begin{equation}
\Gamma^\alpha{}_{\mu\nu} = \hat{\Gamma}^\alpha{}_{\mu\nu} + L^\alpha{}_{\mu\nu}\,,
\label{eq:conn_decomposition_L}
\end{equation}
where $\hat{\Gamma}^\alpha{}_{\mu\nu}$ is the Levi-Civita connection of the metric $g_{\mu\nu}$ and
$L^\alpha{}_{\mu\nu}$ is the disformation tensor, defined in terms of nonmetricity as
\begin{equation}
L^\alpha{}_{\mu\nu} \equiv - \frac{1}{2} g^{\alpha\beta} \Big(
Q_{\mu\nu\beta} + Q_{\nu\mu\beta} - Q_{\beta\mu\nu} \Big)\,.
\label{eq:disformation_L}
\end{equation}
The disformation is symmetric in its lower two indices, $L^\alpha{}_{\mu\nu} = L^\alpha{}_{\nu\mu}$,
and encodes the deviation of the affine connection from Levi-Civita due solely to nonmetricity.
This decomposition ensures that the connection satisfies the torsionless and curvatureless conditions
while retaining general covariance.

The STG formulation also admits a superpotential $P^\alpha{}_{\mu\nu}$, constructed from nonmetricity invariants. Explicitly, for a general quadratic nonmetricity action, the superpotential can be written as \cite{BeltranJimenez:2019esp}
\begin{equation}
P^\alpha{}_{\mu\nu} = \frac{1}{4} \big( -Q^\alpha{}_{\mu\nu} + 2 Q_{(\mu\nu)}{}^\alpha - Q^\alpha g_{\mu\nu} - \tilde{Q}^\alpha g_{\mu\nu} \big)\,,
\end{equation}
where $Q_\alpha = Q_\alpha{}^\mu{}_\mu$ and $\tilde{Q}_\alpha = Q^\mu{}_{\alpha\mu}$ are the two independent traces of nonmetricity.
This tensor enters the field equations once the gravitational Lagrangian is specified.

\subsection{Light propagation}
\label{sec:light}

The propagation of electromagnetic radiation in a general metric-affine spacetime can be consistently analyzed within the geometrical-optics approximation \cite{Perlick:2000, Hehl:2003}.
In this section, we summarize the kinematical properties of light propagation that are relevant for cosmological distance relations, with particular emphasis on the role of nonmetricity and on the conservation of photon number.

Following \cite{Santana:2017zvy}, the electromagnetic four-potential can be expanded as
\begin{equation}
\mathcal A_\mu = \Re\left\{\left[\alpha_\mu + \epsilon\, \beta_\mu + \mathcal O(\epsilon^2)\right]e^{iS/\epsilon}\right\},
\label{eq:EM_potential}
\end{equation}
where \(S\) is the rapidly varying eikonal function, $\epsilon \ll 1$ is a dimensionless expansion parameter, and the wave covector is defined as $k_\mu \equiv \partial_\mu S$.

At leading order in \(\epsilon\), the Maxwell equations imply the null condition
\begin{equation}
k^\mu k_\mu = 0\,,
\end{equation}
so that electromagnetic waves propagate along metric null directions,
independently of curvature, torsion, or nonmetricity.

At next-to-leading order, the wave covector is transported along the rays according to the standard geodesic equation of the metric connection:
\begin{equation}
k^\nu \hat{\nabla}_\nu k^\mu = 0\,,
\end{equation}
where \(\hat{\nabla}_\nu\) denotes the Levi-Civita covariant derivative. 
Consequently, light rays are metric null geodesics, although they are not necessarily autoparallels of the full affine connection. 
This ensures that the causal structure and the kinematics of null congruences can be treated using standard Riemannian tools.

The evolution of the wave amplitude arises at the same next-to-leading order in \(\epsilon\). 
For minimal electromagnetic coupling, it satisfies the standard transport equation
\begin{equation}
\hat{\nabla}_\mu \left(\alpha^2 k^\mu \right) = 0\,,
\label{eq:photon_conservation}
\end{equation}
where \(\alpha^2 \equiv \alpha_\mu \alpha^\mu\) is the squared amplitude of the
electromagnetic potential. 
Eq.~\eqref{eq:photon_conservation} expresses the conservation of the photon-number current along metric null geodesics and follows exclusively from minimal coupling assumptions, independent of the gravitational field equations. 

The above results establish that, even in the presence of spacetime nonmetricity, light propagates along metric null geodesics and its flux is conserved. 
Accordingly, any deviation from the standard DDR cannot be attributed to modified light trajectories or photon nonconservation. 
Instead, such deviations must arise from geometric effects of nonmetricity on the evolution of infinitesimal light beams, which will be analyzed explicitly in the following sections.

\section{Distance reciprocity in metric-affine spacetimes}
\label{sec:metric-affine}

We consider a congruence of null curves with a tangent vector $k^\mu = dx^\mu/d\lambda$, where $\lambda$ is an affine parameter with respect to the Levi-Civita connection associated with the metric $g_{\mu\nu}$.
As shown in Sec.~\ref{sec:light}, in the case of minimally coupled electromagnetism, the wave covector satisfies
\begin{equation}
k^\mu k_\mu = 0\,,
\quad
k^\nu \hat{\nabla}_\nu k^\mu = 0\,,
\label{eq:null_geo}
\end{equation}
so that photon trajectories are null geodesics of the metric
connection.

An infinitesimal light beam around a fiducial null geodesic is described
by two linearly independent connecting vectors $X^\mu$ and $Y^\mu$
linking neighboring geodesics.
They are chosen to lie in the screen space orthogonal to $k^\mu$ and to
the observer four-velocity $u^\mu$,
\begin{equation}
k_\mu X^\mu = k_\mu Y^\mu = 0\,,
\quad
u_\mu X^\mu = u_\mu Y^\mu = 0\,.
\label{eq:screen}
\end{equation}

Since neighboring photons follow neighboring metric geodesics, the
connecting vectors are transported using the Levi-Civita connection.
Their evolution is governed by the geodesic deviation equation
\begin{equation}
\frac{D^2 X^\mu}{d\lambda^2}=\hat R^\mu{}_{\nu\rho\sigma}k^\nu k^\rho X^\sigma\,,
\label{eq:dev}
\end{equation}
and analogously for $Y^\mu$. Here, $D/d\lambda\equiv k^\mu\hat\nabla_\mu$ denotes the covariant derivative along the null geodesic with respect to the Levi-Civita connection.

We then introduce the antisymmetric scalar
\begin{equation}
\Xi \equiv
Y_\mu \frac{D X^\mu}{d\lambda}-
X_\mu \frac{D Y^\mu}{d\lambda}\,.
\label{eq:Xi_def}
\end{equation}
Taking a derivative along the geodesic yields
\begin{align}
\frac{d\Xi}{d\lambda}
&=
Y_\mu \frac{D^2 X^\mu}{d\lambda^2}
-
X_\mu \frac{D^2 Y^\mu}{d\lambda^2}
+
\frac{D Y_\mu}{d\lambda}\frac{D X^\mu}{d\lambda}
-
\frac{D X_\mu}{d\lambda}\frac{D Y^\mu}{d\lambda}.
\end{align}
The last two terms cancel identically due to antisymmetry.
Substituting Eq.~\eqref{eq:dev} gives
\begin{equation}
\frac{d\Xi}{d\lambda}
=
\hat R_{\mu\nu\rho\sigma}
k^\nu k^\rho
\left(
Y^\mu X^\sigma - X^\mu Y^\sigma
\right).
\end{equation}
The tensor in parentheses is antisymmetric under
$\mu\leftrightarrow\sigma$, while the Riemann tensor is symmetric under
the same exchange when contracted with $k^\nu k^\rho$.
Therefore,
\begin{equation}
\frac{d\Xi}{d\lambda}=0\,.
\label{eq:Xi_cons}
\end{equation}

For an infinitesimal beam spanned by $X^\mu$ and $Y^\mu$, the physical
cross-sectional area is $\Sigma \propto |\Xi|$.
The angular diameter distance is defined as
\begin{equation}
D_A^2 = \frac{\Sigma}{\delta\Omega}\,,
\end{equation}
where $\delta\Omega$ is the solid angle subtended at the source.

Let $D_S$ and $D_R$ denote the area distances evaluated at the source and
at the observer.
Since $\Xi$ is conserved,
\begin{equation}
D_S = D_R\,.
\label{eq:DSDR}
\end{equation}
Using the definition of redshift,
\begin{equation}
1+z = \frac{(k_\mu u^\mu)_S}{(k_\mu u^\mu)_R}\,,
\label{eq:redshift}
\end{equation}
and identifying the angular diameter distance with the area distance at
the source, $D_A \equiv D_S$, we obtain
\begin{equation}
D_A = (1+z)\, D_R\,,
\end{equation}
which coincides with the standard Etherington relation.

This result confirms that in STG gravity, as long as the electromagnetic sector remains minimally coupled, the nonmetricity does not introduce any anomalous scaling in the distance-redshift relations. Therefore, any observationally detected deviation from the DDR would be a smoking gun for nonminimal coupling or photon number non-conservation.

\section{Nonminimal electromagnetic coupling in $f(Q)$ gravity}
\label{sec:f(Q)}

We now extend the analysis by considering a dynamical STG theory with a nonminimal coupling between electromagnetism and nonmetricity.
The total action reads \cite{Heisenberg:2023lru}
\begin{equation}
\mathcal S_{\rm tot} = \frac{1}{2}\int d^4x \sqrt{-g}\, f(Q) + \mathcal S_{\rm m} + \mathcal S_{\rm EM}\,,
\label{action_total}
\end{equation}
where $g$ is the determinant of the metric $g_{\mu\nu}$ and $f(Q)$ is a general function of the nonmetricity scalar $Q$.
Here, $\mathcal S_{\rm m} = \int d^4x \sqrt{-g}\, \mathcal L_{\rm m}$ accounts for standard matter, while
\begin{equation}
\mathcal S_{\rm EM} = - \frac{1}{4} \int d^4x \sqrt{-g}\, \mathcal I(Q) F_{\mu\nu} F^{\mu\nu}
\end{equation}
represents the electromagnetic sector with a nonminimal coupling $\mathcal I(Q)$ to nonmetricity.  
The field strength is defined as
\begin{equation}
F_{\mu\nu} \equiv \hat\nabla_\mu \mathcal A_\nu - \hat\nabla_\nu \mathcal A_\mu\,,
\end{equation}
and the nonmetricity scalar is expressed in terms of the disformation tensor as (cf. Eq.~\eqref{eq:disformation_L})
\begin{equation}
Q \equiv - g^{\mu\nu} \Big( L^\sigma{}_{\rho\mu} L^\rho{}_{\nu\sigma} - L^\sigma{}_{\rho\sigma} L^\rho{}_{\mu\nu} \Big)\,.
\end{equation}

Varying the action~\eqref{action_total} with respect to $g_{\mu\nu}$ gives the gravitational field equations:
\begin{align}
&\frac{2}{\sqrt{-g}} \hat\nabla_\sigma \big( \sqrt{-g} \tilde f_Q P^\sigma{}_{\mu\nu} \big)
+ \tilde f_Q \left( P_{\mu\sigma\rho} Q_\nu{}^{\sigma\rho} - 2 Q_{\sigma\rho\mu} P^{\sigma\rho}{}_\nu \right) \notag \\
&+ \frac{1}{2} g_{\mu\nu} f = T_{\mu\nu}^{(\rm m)} + \mathcal I(Q) T_{\mu\nu}^{(\rm EM)}\,,
\label{eq:fQ_field_eq}
\end{align}
where $\tilde f_Q \equiv f_Q - \frac{1}{4} \mathcal I_Q F^{\mu\nu} F_{\mu\nu}$, with $f_Q \equiv df/dQ$ and $\mathcal I_Q \equiv d\mathcal I/dQ$.  
The energy-momentum tensors are
\begin{align}
T_{\mu\nu}^{(\rm m)} &= -\frac{2}{\sqrt{-g}} \frac{\delta(\sqrt{-g}\, \mathcal L_m)}{\delta g^{\mu\nu}}, \\
T_{\mu\nu}^{(\rm EM)} &= F_{\mu\sigma} F_\nu{}^\sigma - \frac{1}{4} g_{\mu\nu} F_{\sigma\rho} F^{\sigma\rho}.
\end{align}

Furthermore, varying $\mathcal S_{\rm EM}$ with respect to $\mathcal A_\mu$ gives
\begin{equation}
\hat\nabla_\mu \big[ \mathcal I(Q) F^{\mu\nu} \big] = 0\,,
\label{eq:Maxwell_modified}
\end{equation}
which represent the modified Maxwell equations in the presence of the nonminimal coupling.  

A fundamental requirement for the consistency of the electromagnetic sector is the preservation of the local ${\rm U}(1)$ gauge symmetry. Since the field strength tensor is defined through the exterior derivative as $F=d\mathcal{A}$, its components are structurally independent of the affine connection and the nonmetricity tensor. Consequently, the gauge transformation $\mathcal{A}_\mu \to \mathcal{A}_\mu + \partial_\mu \Lambda$ leaves the field strength invariant. 
From a geometrical perspective, the ${\rm U}(1)$ gauge field represents a connection on a principal fiber bundle over the spacetime manifold \cite{Nakahara:2003nw}. The coupling function $\mathcal I(Q)$ depends only on the scalar nonmetricity invariant and, hence, does not introduce any direct interaction between the ${\rm U}(1)$ connection and the affine ${\rm GL}(4,\mathbb R)$ connection associated to the spacetime geometry. As a result, the gauge invariance is not affected by the choice of the holonomic connection.
As will be discussed in Sec.~\ref{sec:dynamical_violation}, in the coincident gauge, $Q$ transforms as a proper scalar under general coordinate transformations. This ensures that the coupling $\mathcal{I}(Q)F_{\mu\nu}F^{\mu\nu}$ preserves local Lorentz invariance and does not affect the ${\rm U(1})$ bundle structure over the ${\rm GL}(4,\mathbb{R})$ manifold \cite{BeltranJimenez:2017tkd}. The modified Maxwell equations \eqref{eq:Maxwell_modified} remain thus consistent with the conservation of the electromagnetic current.

\subsection{Propagation of electromagnetic waves}
\label{sec:f(Q)A}

We shall now study the propagation of electromagnetic waves within the framework of the model~\eqref{action_total}.

Substituting the ansatz~\eqref{eq:EM_potential} into the modified Maxwell equations~\eqref{eq:Maxwell_modified}, one finds at leading order
\begin{equation}
k_\mu k^\mu = 0\,,
\end{equation}
showing that photons propagate along null geodesics of the background metric.
Hence, the nonminimal coupling $\mathcal I(Q)$ does not affect photon trajectories at this order.

At next-to-leading order, one has
\begin{equation}
 \hat\nabla_\mu \Big[ \mathcal{I}(Q) \big(\hat\nabla^\mu \alpha^\nu - \hat\nabla^\nu \alpha^\mu + i (k^\mu \beta^\nu - k^\nu \beta^\mu)\big) \Big] = 0\,.
 \label{eq:NLO_1}
\end{equation}
Multiplying by $\alpha_\nu$ and taking only the real part\footnote{The real part of Eq.~\eqref{eq:NLO_1} provides the dynamics of the amplitude of the electromagnetic potential, while the imaginary part governs the polarization modes.}  of the above relation leads to
\begin{align}
    &\alpha_\nu\hat\nabla_\mu\big[\mathcal{I}(\hat\nabla^\mu\alpha^\nu-\hat\nabla^\nu\alpha^\mu)\big]
    =\hat\nabla_\mu\left[\mathcal{I}(\alpha_\nu\hat\nabla^\mu\alpha^\nu-\alpha_\nu \hat\nabla^\nu\alpha^\mu)\right]\notag \\
    &+\mathcal{I}\big[(\hat\nabla_\mu\alpha_\nu)(\hat\nabla^\nu\alpha^\mu)
    -(\hat\nabla_\mu\alpha_\nu)(\hat\nabla^\mu\alpha^\nu)\big]=0\,.
    \label{eq:starting_transport}
\end{align}
We note that in the geometric-optics approximation, 
$|\hat\nabla\alpha|\ll|k\alpha|$, we can neglect the terms 
$(\hat\nabla\alpha)^2$ in the second line of Eq.~\eqref{eq:starting_transport}. 
Hence, we find
\begin{equation}
    \hat\nabla_\mu\big[\mathcal{I}(\alpha_\nu\hat\nabla^\mu\alpha^\nu
    -\alpha_\nu\hat\nabla^\nu\alpha^\mu)\big]=0\,.
\end{equation}
Using the identity $\alpha^2\equiv\alpha_\mu\alpha^\mu$, we recast the above expression into
\begin{equation}
    \hat\nabla_\mu\Big[\mathcal{I}\Big(\frac{1}{2}\hat\nabla^\mu(\alpha^2)
    -\alpha_\nu\hat\nabla^\nu\alpha^\mu\Big)\Big]=0\,.
    \label{eq:NLO_2}
\end{equation}
Projecting onto the transverse subspace and using the leading-order equation, one can show that \cite{Dolan:2017zgu}
\begin{equation}
    \frac{1}{2}\hat\nabla^\mu(\alpha^2)-\alpha_\nu\hat\nabla^\nu\alpha^\mu
    \simeq \frac{1}{2}\alpha^2k^\mu\,,
    \label{eq:Dolan}
\end{equation}
where higher-order terms have been neglected. 

Inserting Eq.~\eqref{eq:Dolan} into Eq.~\eqref{eq:NLO_2}, we finally obtain the transport equation for the electromagnetic amplitude:
\begin{equation}
\hat\nabla_\mu \big[ \mathcal I(Q) \alpha^2 k^\mu \big] = 0\,.
\label{eq:NLO_final}
\end{equation}
Hence, the quantity conserved along the null congruence is not the standard photon current, but a modified current weighted by $\mathcal I(Q)$.

Introducing the photon-number current $J^{\mu} \equiv \mathcal I(Q)\,\alpha^{2}k^{\mu}$, Eq.~\eqref{eq:NLO_final} takes the more compact form
\begin{equation}
\hat\nabla_{\mu}J^{\mu}= 0\,.
\label{eq:transport}
\end{equation}
This represents a genuinely dynamical modification of light propagation, consequently leading to a deviation from the standard DDR.

\subsection{Dynamical violation of the DDR}
\label{sec:dynamical_violation}

To explicitly compute the dynamical violation of the DDR, we specialize to the spatially flat Friedmann-Lema\^itre-Robertson-Walker (FLRW) metric
\begin{equation}
ds^2 = -dt^2 + a^2(t)\delta_{ij}dx^i dx^j \,,
\label{eq:FLRW_metric}
\end{equation}
where $a(t)$ is the scale factor normalized to unity at the present time.
We also adopt a particularly convenient gauge choice, the \emph{coincident gauge}, in which the affine connection vanishes globally, $\Gamma^\alpha{}_{\mu\nu} = 0$ \cite{BeltranJimenez:2019tme,DAmbrosio:2021pnd}.
Consequently, the gravitational degrees of freedom are entirely encoded in the metric,
with the affine structure trivialized. 
In this gauge, the nonmetricity tensor reduces to a simple partial derivative of the metric,
\begin{equation}
Q_{\alpha\mu\nu} = \partial_\alpha g_{\mu\nu}\,, \quad 
Q_{tij} = 2 H g_{ij}\,, 
\end{equation}
with $H \equiv \dot a/a$ being the Hubble expansion parameter.

We consider a narrow bundle of null rays with cross-sectional area $\Sigma$ and unit normal $n_\mu$. The number of photons crossing per unit affine parameter is given by the photon-number current~\eqref{eq:transport}:
\begin{equation}
\mathcal N \propto J^\mu n_\mu \Sigma = \mathcal I(Q) \, \alpha^2 \, (k^\mu n_\mu) \, \Sigma\,.
\label{eq:photon_current}
\end{equation}
Assuming that the source is intrinsically standard, such that the number of emitted photons remains constant, any observed deviation in the flux must be attributed to the evolution of $\mathcal I(Q)$ along the line of sight. 

Integrating Eq.~\eqref{eq:photon_current} along the ray from the source to the observer, we obtain
\begin{equation}
\mathcal I(Q_S) \, \alpha_S^2 \, (k^\mu n_\mu)_S \, \Sigma_S = 
\mathcal I(Q_R) \, \alpha_R^2 \, (k^\mu n_\mu)_R \, \Sigma_R\,.
\end{equation}
Since photons still propagate along null metric geodesics, their energy redshift and the cosmological time-dilation effects retain their standard form. Consequently, the geometrical derivation of Etherington's reciprocity relation remains unchanged, and the observed flux still exhibits the usual $(1+z)^{-4}$ suppression. The only modification arises from the non-conservation of the photon number induced by the nonminimal coupling function $\mathcal I(Q)$. The observed flux can thus be written as
\begin{equation}
\mathcal F_{\rm obs}=\frac{L}{4\pi D_A^2}\,\frac{\mathcal I(Q_S)}{\mathcal I(Q_R)}\,\frac{1}{(1+z)^4}\,.
\label{eq:flux_modified}
\end{equation}
Comparing with the standard luminosity-flux relation $\mathcal F_{\rm obs} = L/(4 \pi D_L^2)$, we find the modified luminosity distance to be
\begin{equation}
D_L=(1+z)^2 D_A
\sqrt{\frac{\mathcal I(Q_R)}{\mathcal I(Q_S)}}.
\label{eq:DDR_dynamical_violation}
\end{equation}
It is worth noting that in this scenario, $D_A$ coincides with the standard metric definition, as the beam geometry is governed by the metric connection, while the DDR is violated dynamically by $\mathcal I(Q)$.

While Eq.~\eqref{eq:DDR_dynamical_violation} is formally analogous to results found in models involving axion-like particles or dilaton-electromagnetism~\cite{Hees:2014lfa,Minazzoli:2013ara}, our derivation provides a distinct geometric interpretation. Unlike models that rely on the existence of exotic particles interacting with the Maxwell sector, the violation of the Etherington relation here is a direct consequence of the nonmetric nature of the gravitational background.
Specifically, although photons still follow null geodesics, the nonminimal coupling $\mathcal I(Q)$ modifies the conservation of the adiabatic invariant (photon number) along the light-cone.
This result establishes that the DDR is a powerful tool to distinguish between standard Riemannian gravity and STG extensions, where the interaction between light and nonmetricity fundamentally alters the flux-distance scaling.

In the coincidence gauge, where $Q = 6H^2$, and for a receiver located at the present time $(z=0)$, Eq.~\eqref{eq:DDR_dynamical_violation} becomes
\begin{equation}
D_L(z)=(1+z)^2 D_A(z)\,\eta(z)\,,
\label{eq:DDR_coicident}
\end{equation}
where the DDR parameter is defined as
\begin{equation}
    \eta(z)\equiv
\frac{D_L(z)}
{(1+z)^2D_A(z)} = \sqrt{\frac{\mathcal I\left(H_0^2\right)}{\mathcal I\left(H^2(z)\right)}}\,.
    \label{eq:eta}
\end{equation}
Standard Etherington's relation is recovered for $\eta(z)=1$.

By relating the DDR violation directly to the Hubble expansion rate, Eq.~\eqref{eq:eta} provides a practical framework for observational tests. Remarkably, once the functional form of the nonminimal coupling $\mathcal I(Q)$ in the action~\eqref{action_total} is specified, Eq.~\eqref{eq:eta} can be used to constrain the viability of a given cosmological model using distance measurements, thereby probing possible departures from GR.

It is worth emphasizing that the coincident gauge represents a self-consistent choice imposed by the STG geometry in a spatially flat FLRW background. Potential concerns regarding boundary terms are addressed by the fact that the variational principle in $f(Q)$ gravity is well-posed in the coincident gauge, as the boundary contributions are consistent with those of the equivalent GR action \cite{BeltranJimenez:2018vdo,DAmbrosio:2021pnd}. Consequently, Eq.~\eqref{eq:eta} should not be interpreted as a mere reparameterization of $H(z)$. Instead, it establishes a direct, nontrivial correlation between the expansion history and the modifications of $D_L(z)$. 

Standard observational tests of the DDR in the literature are performed by assuming phenomenological parametrizations of the redshift dependence, with specific functional forms fitted to data. In contrast, our analysis shows that the DDR violation is directly linked to the Hubble expansion rate via the functional form of the nonminimal coupling $\mathcal I(Q(H))$. This direct dependence allows one to probe the underlying gravitational dynamics without introducing arbitrary redshift parametrizations, providing a more physically grounded connection between theory and observation. Such a formulation therefore offers a potentially distinctive observational signature that can help to break degeneracies among different modified gravity scenarios, enabling a direct test of non-Riemannian effects on cosmological distance measures.

A further aspect worth mentioning is that the same nonminimal coupling responsible for the DDR violation also induces a variation of the fine-structure constant. Indeed, in theories where the electromagnetic field is coupled to a function of a scalar field, $h(\phi)$, the effective fine-structure constant satisfies $\alpha_{\rm em}\propto h^{-1}(\phi)$ \cite{Damour:1994zq,Damour:2002mi}. In the present framework, the role of $h(\phi)$ is played by the function $\mathcal I(Q)$. Hence, we can write
\begin{equation}
\alpha_{\rm em}(z)\propto \mathcal I^{-1}(Q(z))\,.
\label{eq:DDR_fine-structure}
\end{equation}
Combining this with Eq.~\eqref{eq:eta}, it follows that \cite{Hees:2014lfa}
\begin{equation}
\eta^2(z)=\frac{\alpha_{\rm em}(z)}{\alpha_{\rm em,0}}\,.
\label{eq:DDR_fine-structure_2}
\end{equation}
Thus, variations of $\alpha_{\rm em}$ and violations of the DDR are not independent effects but constitute complementary probes of the same nonminimal coupling sector, providing a consistency relation that may be tested with increasing precision by forthcoming cosmological observations.

Current astrophysical and laboratory measurements place stringent constraints on the variation of the fine-structure constant, typically at the level $|\Delta\alpha_{\rm em}/\alpha_{\rm em}|\lesssim10^{-6}-10^{-5}$ over cosmological timescales \cite{Webb:1998cq,Uzan:2010pm,King:2012id,Martins:2017yxk}. These bounds imply that the coupling function $\mathcal I(Q)$ must remain very close to its present value throughout the observable redshift range. As a consequence, the DDR violation predicted by our model is also expected to be small, consistent with current analyses based on galaxy clustering and cosmological probes, which typically constrain departures from Etherington's relation at the few-percent level over a broad redshift range \cite{Avgoustidis:2010ju,Holanda:2015zpz,Liao:2015uzb,Lin:2018qal,Li:2025htp}.

\subsection{Examples of DDR violation}
In the next paragraphs, we shall consider three illustrative examples of electromagnetic couplings leading to a dynamical violation of the DDR according to the results obtained in Sec.~\ref{sec:dynamical_violation}.

\subsubsection{Power-law coupling} 

Let us first consider the following coupling:
\begin{equation}
    \mathcal I(Q)=1+\epsilon\left(\frac{Q}{Q_0}\right)^{-n}\,, \quad n>0\,.
    \label{eq:power_coupling}
    \end{equation}
Here, $\epsilon\ll 1$ is a small dimensionless coupling constant, and $Q_0$ is the current value of the nonmetricity scalar. 
In the coincident gauge, Eq.~\eqref{eq:power_coupling} translates to
\begin{equation}
    \mathcal{I}(H^2)=1+\epsilon\left(\frac{H_0^2}{H^2}\right)^n\,,
\end{equation}
which yields $\mathcal I(H_0^2)=1+\epsilon$ at the current epoch. From the definition of the DDR violation parameter, it follows that $\eta(0)=1$ holds identically, meaning that the standard Etherington relation is exactly restored at present.

Then, to evaluate the behaviour at $z>0$, we substitute the above expressions into Eq.~\eqref{eq:eta}, and perform a first-order expansion in $\epsilon$:
\begin{equation}
    \eta(z)\simeq 1+\frac{\epsilon}{2}\left[1-\left(\frac{H_0}{H(z)}\right)^{2n}\right].
    \label{eq:power}
\end{equation}
Notice that, in the early Universe where $H\gg H_0$, the electromagnetic coupling vanishes, leading to the maximum DDR deviation, $\eta\simeq 1+\epsilon/2$. This ensures that $\eta(z)$ remains sufficiently close to unity throughout the whole cosmic history, thereby satisfying both the DDR and $\alpha_{\rm em}$ observational bounds.

\subsubsection{Exponential coupling}

Alternatively, we can consider an exponential coupling function of the form
\begin{equation}
    \mathcal I(Q)=1+\epsilon\, e^{-Q/Q_0 }\,,
\end{equation}
with $\epsilon\ll 1$. Under the coincident gauge condition, this parametrization reads
\begin{equation}
    \mathcal{I}(H^2)=1+\epsilon\, e^{-(H/H_0)^2},
\end{equation}
which implies $\mathcal{I}(H_0^2) = 1 + \epsilon\, e^{-1}$ at the present era.

Similarly to the previous case, using Eq.~\eqref{eq:eta} and performing a first-order expansion in $\epsilon$, we find
\begin{equation}
    \eta(z)\simeq  1+\frac{\epsilon}{2e}\left[1-\exp\left(1-\frac{H^2(z)}{H_0^2}\right)\right].
    \label{eq:exp}
\end{equation}
In this scenario, for $H\gg H_0$ in the early epoch, the exponential term strongly suppresses the nonminimal coupling, leading to $\eta(z)\simeq 1+\epsilon/(2e)$. This deviation smoothly decreases up to the present time, when $\eta(0)=1$, consistent with the lack of significant DDR violations from current astrophysical and cosmological observations.

\subsubsection{Logarithmic coupling} 

Finally, we can explore a logarithmic coupling function parameterized as
\begin{equation}
    \mathcal I(Q)=1+\epsilon \ln \left(1+\frac{Q_0}{Q}\right),
\end{equation}
where $\epsilon \ll 1$. In the coincident gauge, this becomes
\begin{equation}
    \mathcal{I}(H^2)=1+\epsilon \ln \left(1+\frac{H_0^2}{H^2}\right),
\end{equation}
which implies $\mathcal{I}(H_0^2) = 1 + \epsilon \ln 2$ at the present epoch. As in the previous cases, the standard DDR is exactly preserved today, since $\eta(0)=1$ holds identically. 

To describe the evolutionary behaviour at higher redshifts, we expand Eq.~\eqref{eq:eta} to the first order in $\epsilon$, obtaining
\begin{equation}
    \eta(z)\simeq 1+\frac{\epsilon}{2} \ln \left[\frac{2 H^2(z)}{H^2(z) + H_0^2} \right].
    \label{eq:log}
\end{equation}
At very high redshifts ($H \gg H_0$), the logarithmic term approaches zero, and the DDR deviation reaches its maximum value given by $\eta \simeq 1 + \epsilon \ln \sqrt{2}$. Also in this case, the function $\eta(z)$ remains always very close to unity, as indicated by observations.

\begin{figure}
    \centering
    \includegraphics[width=3.3in]{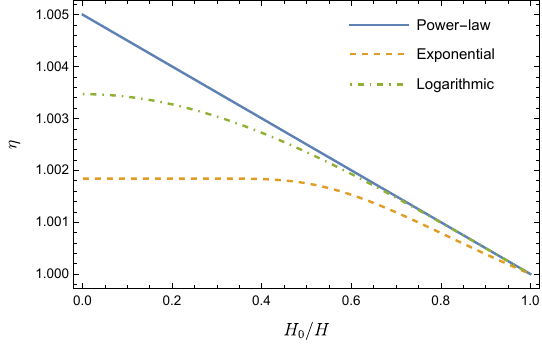}
    \caption{Cosmic evolution of the DDR violation parameter for different electromagnetic couplings: power-law ($n=1$) (cf. Eq.~\eqref{eq:power}), exponential (cf. Eq.~\eqref{eq:exp}), and logarithmic (cf. Eq.~\eqref{eq:log}) functional forms for $\epsilon=0.01$.}
    \label{fig:eta}
\end{figure}

In Fig.~\ref{fig:eta}, we display the cosmic evolution of the DDR parameter $\eta(z)$ from the early Universe to the present epoch, arising from the different parametrizations of the electromagnetic coupling function considered above. In all cases, $\eta(z)$ remains very close to unity over the entire redshift range. In particular, the deviations from the standard Etherington relation satisfy $\eta-1< 0.5\%$ for $\epsilon=10^{-2}$, and $\eta-1< 0.05\%$ for $\epsilon=10^{-3}$.

\section{Final remarks}
\label{sec:final}

In this work, we have investigated the fundamental relations connecting different notions of cosmological distance within the framework of symmetric teleparallel gravity. By analyzing light propagation in a metric-affine spacetime and adopting the geometric optics approximation, we have shown that spacetime nonmetricity, under the assumption of minimal electromagnetic coupling, does not intrinsically alter the standard Etherington relation. 
This result establishes that the kinematics of null geodesics and the conservation of the photon number current in STG are consistent with the standard Riemannian predictions, provided the electromagnetic sector remains conventional.

We have then extended our study to characterize a second, physically distinct source of DDR violation in $f(Q)$ gravity.
We have demonstrated that a nonminimal coupling between electromagnetism and the nonmetricity scalar leads to a genuinely dynamical violation of the DDR. In this scenario, the standard photon number current is no longer conserved, as its transport is modulated by the evolution of the nonmetricity scalar along the line of sight. This effect produces a modified luminosity distance that deviates from the standard scaling, offering a concrete mechanism to probe the foundations of non-Riemannian gravity.

For a flat FLRW background in the coincident gauge, we have derived a practical relation that links the DDR violation to the Hubble expansion rate. This formula provides a unified framework to disentangle geometric and dynamical effects, which is crucial for the correct interpretation of cosmological observations.
An interesting observational signature of the framework is the direct link between DDR violations and variations of the fine-structure constant. The combination of distance measurements with high-precision determinations of $\alpha_{\rm em}$ from future surveys may therefore provide a powerful test of the nonmetricity-induced coupling.

To demonstrate the phenomenological viability of our framework, we have analyzed three distinct functional forms for the nonminimal electromagnetic coupling. Crucially, all these scenarios share the fundamental property of exactly restoring the standard Etherington relation at the present epoch. At the same time, they predict a well-behaved, maximum deviation in the deep past. We have shown that the resulting deviations from the standard DDR remain strictly within sub-percent levels throughout the entire cosmic history, without entering into conflict with both DDR tests and $\alpha_{\rm em}$ variations.

From an observational perspective, our findings open promising avenues for future applications. 
In contrast with standard approaches in the literature, which rely on phenomenological parametrizations of the DDR as functions of redshift, the modified DDR derived here is directly tied to the Hubble expansion rate through the nonminimal coupling.
This direct connection grounds the analysis in the underlying gravitational dynamics, avoiding arbitrary assumptions about redshift dependence and offering a more physically motivated avenue to constrain modified gravity effects. Such an approach could reveal distinctive signatures of nonmetricity that are not accessible through conventional parametrized tests.

The obtained modified DDR can be directly employed in analyses of type Ia supernovae, baryon acoustic oscillations, and cosmic microwave background data.
Furthermore, the comparison between electromagnetic and gravitational-wave luminosity distances in multi-messenger astronomy could provide even more stringent constraints on the nonminimal couplings predicted by STG theories. In conclusion, our results offer a comprehensive framework to assess the viability of $f(Q)$ gravity as a compelling alternative to standard GR.

\acknowledgments
The author thanks the anonymous referee for the helpful and valuable comments.
The author acknowledges financial support from the Istituto Nazionale di Fisica Nucleare (INFN), Sezione di Roma 1, \textit{esperimento} Euclid. This paper is based upon work from COST Action CA21136 -- Addressing observational tensions in cosmology with systematics and fundamental physics (CosmoVerse), supported by European Cooperation in Science and Technology.

\bibliography{references}

\end{document}